\documentclass[
11pt,
prd,
showpacs,
nofootinbib,
oneside,
hidelinks,
a4paper,
english,
aps
]{revtex4-2}

\usepackage[utf8]{inputenc}
\usepackage[T1]{fontenc}

\usepackage{amsmath}
\usepackage{amssymb}
\usepackage{bm}
\usepackage{braket}

\usepackage{graphicx}
\usepackage[dvipsnames]{xcolor}
\usepackage{hyperref}

\hypersetup{
	colorlinks=true,
	linkcolor=red,
	citecolor=blue,
	urlcolor=blue
}

\begin{document}
	
	
	%
	%
	
	\title{Enthalpy-Based Thermal Response and Its Exact Relation to the Speed of Sound in Finite-Temperature QCD}
	
	\author{S. D. Campos}\email{ sergiodc@ufscar.br}
	\affiliation{Applied Mathematics Laboratory-DFQM/CCTS, Federal University of S\~ao Carlos, Sorocaba, S\~ao Paulo CEP 18052780, Brazil}
	
	\begin{abstract}
		We quantify the logarithmic thermal variation of the normalized enthalpy density in finite-temperature Quantum Chromodynamics using a dimensionless response thermal function, $\mathcal{H}(T)$. We establish an exact identity that links $\mathcal{H}(T)$ to the speed of sound, $c_s^2(T)$. Using continuum-extrapolated lattice Quantum Chromodynamics, Monte Carlo uncertainty propagation, and cubic spline interpolation, we extract a stable peak at $T_{\text{peak}} \approx 153.6\text{ MeV}$ ($\mathcal{H}_{\text{peak}} \approx 6.24$), which remains robust under changes in the smoothing parameter $s$. By contrast, $\mathcal{H}(T) = 0$ for the MIT Bag Model despite its non-vanishing trace anomaly. We highlight $\mathcal{H}(T)$ as an effective diagnostic of the QCD crossover and discuss its limitations for universal critical scaling at zero chemical potential.
	\end{abstract}


	
	\maketitle

	\section{Introduction}\label{sec:intro}
	
	The thermodynamic phase structure of Quantum Chromodynamics (QCD), in particular, the location and nature of phase boundaries in the QCD phase diagram, the possible existence of a critical point, and the transition between hadronic and partonic matter, remains a central problem in high-energy nuclear physics
	\cite{borsanyi2025phase, koch2025exploring}. Although perturbative methods provide a reliable description of hard processes at large momentum transfer,
	the transition from a hadronic medium to a deconfined quark-gluon plasma
	(QGP) is governed by non-perturbative dynamics \cite{ratti2018} whose dynamics affect the QCD equation of state (EOS) in the temperature range relevant to the crossover at vanishing or small baryon chemical potential.
	
	A usual measure of the departure from conformal behavior is the trace
	of the energy-momentum tensor \cite{bazavov2009,petreczky2012}. This trace anomaly (also known as the interaction measure) provides important information about the temperature dependence of the EOS and the breaking of scale invariance in QCD \cite{bazavov2019, FuPawlowski2020}. However, this is not the only possible thermodynamic diagnostic for the crossover, as several other observables, such as chiral susceptibility and strangeness fluctuations, have been widely used to characterize the transition \cite{bazavov2012,borsanyi2010}. In particular, the enthalpy density, defined roughly as the sum of the energy density and pressure, directly characterizes the thermodynamic weight carried by the medium and may offer a complementary description of the rapid change in the active degrees of freedom.
	
	Thermodynamic response functions are especially useful in regions where the
	EOS varies rapidly with temperature or chemical potential
	\cite{Pelissetto2002, Dupuis2021}. In the vicinity of a true phase transition, such responses can exhibit nonanalytic behavior in the thermodynamic limit \cite{wilson1974}. At physical quark masses and vanishing baryon chemical potential, finite-temperature QCD is generally expected to undergo a smooth
	crossover rather than a singular phase transition \cite{aoki.2006,bazavov2014}. Consequently, observables constructed from the EOS should be interpreted as finite indicators of rapid thermodynamic variation, unless a nonanalytic critical point is independently established.
	
	In this work, we introduce a dimensionless, enthalpy-based thermal response (hereafter, one writes only thermal response) that measures the logarithmic variation of enthalpy density with temperature after removing the trivial conformal scaling. The construction is written directly in terms of scalar thermodynamic quantities derived from the QCD EOS, ensuring thermodynamic consistency without arbitrary weighting coefficients or auxiliary tensorial structures, relying only on the fundamental relations among pressure, energy density, and enthalpy \cite{borsanyi2010,bazavov2014}. The corresponding temperature-dependent weights are ultimately determined by the relative contributions of the energy density and pressure to the enthalpy. In this formulation, the proposed thermal response quantifies the residual temperature dependence of the normalized enthalpy and therefore serves as an EOS diagnostic of non-conformal thermodynamic behavior.
	
	We examine the properties of this thermal response using both analytical equations of state and continuum-extrapolated lattice-QCD results \cite{borsanyi2010}. In the MIT Bag Model \cite{chodos1974,shuryak1980}, a useful limiting case is that temperature-independent model parameters imply that the vacuum contribution cancels out in the enthalpy density. Consequently, the thermal response does not reproduce the model’s nonzero trace anomaly. This distinction is essential: the proposed thermal response and the trace anomaly are separate thermodynamic observables and should not be unified.
	
	The lattice-QCD analysis is used to determine whether the thermal
	response develops a localized enhancement in the crossover region, which is interpreted as evidence of a rapid change in the normalized
	enthalpy, rather than as a direct signal of a singularity, a universal
	critical susceptibility, or a transport coefficient. We compare its behavior
	with independent EOS observables, including the interaction
	measure and the speed of sound, and discuss the role of interpolation,
	continuum extrapolation, and uncertainty propagation in its numerical
	evaluation.
	
	The paper is organized as follows. Section~\ref{sec:enthalpy} introduces the dimensionless thermal response and derives its representation in terms of thermodynamic quantities obtained from the EOS, including its exact relation to the interaction measure and the squared speed of sound. Section~\ref{sec:ref_eos} discusses analytical reference equations of state, including the MIT Bag Model and the single-pion reference. Section~\ref{sec:lattice} presents continuum-extrapolated lattice-QCD EOS data and describes the numerical procedure, including uncertainty propagation, spline interpolation, and stability tests. Section~\ref{sec:resultados} presents the lattice results for the normalized enthalpy and the enthalpy response, including the determination and stability analysis of its maximum in the crossover region, as well as comparisons with the normalized interaction measure and the squared speed of sound. The consistency of these observables with the relevant thermodynamic identity is also examined. Section~\ref{sec:limitations} summarizes the limitations of the analysis, including its dependence on the chosen EOS and interpolation procedure, as well as the lack of a direct determination of universal critical behavior. Finally, Section~\ref{sec:final_remarks} summarizes the main conclusions and outlines directions for future work.

	\section{Enthalpy-Based Thermal Response}
	\label{sec:enthalpy}
	
	
	The enthalpy density can be defined by the simple relation \cite{callen1985}
	\begin{equation}\label{eq:enthalpy_density}
		w(T)=\epsilon(T)+p(T),
	\end{equation}
	where $\epsilon(T)$ and $p(T)$ are the energy density and pressure,
	respectively. As is well known, the trivial conformal scaling can be removed by writing
	\begin{equation}\label{eq:dimensionless_enthalpy}
		\widetilde{w}(T) \equiv  \frac{w(T)}{T^4} = \frac{\epsilon(T)+p(T)}{T^4}.
	\end{equation}
	
	From the dimensionless result \eqref{eq:dimensionless_enthalpy}, one introduces the main quantity of this work, which is the thermal response
	\begin{equation}\label{eq:enthalpy_response}
		\mathcal{H}(T) \equiv
		\frac{d\ln \widetilde{w}(T/T_i)}{d\ln(T/T_i)}
		= \frac{T}{\widetilde{w}(T/T_i)}\frac{d\widetilde{w}(T/T_i)}{dT}.
	\end{equation}
	where the use of $d/d\ln T$, rather than $d/dT$, ensures that
	$\mathcal{H}(T)$ is dimensionless and, hereafter, one assumes $T_i=1.0$ MeV. Of course, $\mathcal{H}(T)$ vanishes whenever $\widetilde{w}(T)$ is independent of temperature, and, in particular, this occurs in an ideal conformal system, for which $\epsilon=3p\propto T^4$ \cite{aarts2016}. Observe that a nonzero thermal response indicates that the normalized enthalpy varies with temperature, but it does not, by itself, distinguish the microscopic mechanism responsible for that variation. 
	
	To stress the above statement, we have the simple example of a hadronic gas: massive resonances are thermally populated more strongly as $T$ increases, so the effective EOS shifts with temperature, and $\widetilde{w}(T)$ becomes temperature-dependent \cite{karsch2002,braun-munzinger2004}. In the hadronic phase ($T < T_c$), the system can be described by a gas of massive hadronic resonances, whose thermal excitation is exponentially suppressed \cite{karsch2002}. Conversely, for $T > T_c$, hadronic bound states melt into deconfined quarks and gluons, leading to a rapid rise of thermodynamic observables toward the ideal-gas (Stefan--Boltzmann) limit of QCD degrees of freedom \cite{braun-munzinger2004}. In this way, the normalized enthalpy serves as a thermodynamic indicator of the crossover: it changes rapidly but smoothly across the transition region as the relevant degrees of freedom evolve from hadrons to quarks and gluons \cite{borsanyi2010}.
	
	On the other hand, from a field-theoretic perspective, the breaking of conformal symmetry in QCD is encoded in the trace anomaly of the energy-momentum tensor \cite{bazavov2009,petreczky2012,bazavov2014, cheng2008}, written as
	\begin{equation}\label{eq:trace_anomaly}
		\theta_{\mu}^{\mu} = \epsilon - 3p,
	\end{equation}
	and, in a strictly scale-invariant system, the trace vanishes, implying \(\epsilon = 3p\). By contrast, low-temperature QCD lies far from this conformal limit, as confinement and chiral symmetry breaking introduce a large intrinsic scale \cite{wilson1974,S.Weinberg.Phys.Rev.D8.3497.1973}. As a result, the interaction measure is nonzero, making it a useful quantity in lattice QCD. As the temperature increases through the crossover region, these non-perturbative effects vary rapidly, giving rise to a pronounced maximum in the trace anomaly \cite{borsanyi2010,bazavov2014}. Furthermore, in lattice QCD thermodynamics, this observable is central because the pressure and energy density are obtained from the partition function, and the trace anomaly is particularly sensitive to rapid variations in the EOS near the QCD crossover, where it exhibits a pronounced peak in contrast to the conformal limit \cite{karsch2002,borsanyi2010}. 
	
	As is well known, lattice calculations typically determine \((\epsilon - 3p)/T^4\) directly, and the resulting peak is then integrated to obtain \(p(T)\), \(\epsilon(T)\), and other thermodynamic quantities \cite{petreczky2012}. The peak therefore provides both a physical signal of the crossover temperature region and a practical input for constructing the QCD EOS used in hydrodynamic modeling of heavy-ion collisions \cite{aoki.2006,borsanyi2020}.
	
	After this brief digression, we return to the main point and rewrite the equation \eqref{eq:dimensionless_enthalpy} as
	\begin{equation}
		\widetilde{w}(T) = \widetilde{\epsilon}(T)+\widetilde{p}(T),
	\end{equation}
	where $\widetilde{\epsilon}=\epsilon/T^4$ and
	$\widetilde{p}=p/T^4$, resulting the thermal response \eqref{eq:enthalpy_response} can be written as
	\begin{align}\label{eq:weighted_enthalpy_response}
		\mathcal{H}(T) = \frac{\epsilon}{\epsilon+p} \frac{d\ln\widetilde{\epsilon}}{d\ln T} + \frac{p}{\epsilon+p} \frac{d\ln\widetilde{p}}{d\ln T}.
	\end{align}
	
	The coefficients on the right-hand side of the equation \eqref{eq:weighted_enthalpy_response} can be rewritten by means of the auxiliary quantities
	\begin{equation}\label{eq:thermodynamic_weights}
		g(T)=\frac{\epsilon}{\epsilon+p}, \qquad f(T)=\frac{p}{\epsilon+p}, \qquad f(T)+g(T)=1,
	\end{equation}
	implying they are therefore thermodynamic fractions rather than adjustable phenomenological parameters. Then, equation~\eqref{eq:weighted_enthalpy_response} can be written using
	\begin{align}
		\frac{d\ln \widetilde{w}}{d\ln T} = \frac{1}{\epsilon+p}
		\left(\frac{d\epsilon}{d\ln T} + \frac{dp}{d\ln T}\right)-4=
		g(T) \frac{d\ln(\epsilon/T^4)}{d\ln T}
		+f(T)\frac{d\ln(p/T^4)}{d\ln T},
	\end{align}
	providing a thermodynamic justification for the weights in
	equation~\eqref{eq:thermodynamic_weights}. Moreover, this construction removes the arbitrariness associated with assigning constant exponents to $\widetilde{\epsilon}$ and $\widetilde{p}$ by using temperature-dependent weights, which are thermodynamically motivated and thus avoid fixed, {\it ad hoc} power-law choices.
	
	In the approximately conformal high-temperature regime, where
	$\epsilon\simeq 3p$, the weights approach
	\begin{equation}\label{eq:conformal_weights}
		g(T)\longrightarrow \frac{3}{4},
		\qquad
		f(T)\longrightarrow \frac{1}{4},
	\end{equation}
	and near the crossover, however, these weights must generally be retained as temperature-dependent quantities because the ratio $p(T)/\epsilon(T)$ varies with temperature. In the crossover region, their temperature dependence encodes the departure from conformality associated with the trace anomaly and the gradual activation of QCD degrees of freedom. Therefore, keeping \(g(T)\) and \(f(T)\) as dynamical thermodynamic quantities is essential for accurately describing the EOS.
	
	As aforementioned, $\mathcal{H}(T)$ vanishes when the dimensionless enthalpy is temperature-independent, which includes the ideal conformal limit, $\epsilon=3p$, and both $\epsilon/T^4$ and $p/T^4$ are constant. Unlike the trace anomaly, which measures the breaking of conformal symmetry through $\epsilon-3p$, $\mathcal{H}(T)$ characterizes the temperature dependence of the dimensionless enthalpy, and a nonzero value of $\mathcal{H}(T)$ therefore signals a departure from this ideal scaling behavior. In QCD, such a departure receives contributions from the running of the coupling, finite quark masses, and non-perturbative dynamics, being related to, but not identical to, the trace anomaly, written as \cite{bazavov2014,borsanyi2010,petreczky2012}
	\begin{equation}\label{eq:dimensionless_trace_anomaly}
		\frac{\Theta(T)}{T^4} \equiv \frac{\epsilon(T)-3p(T)}{T^4}.
	\end{equation}
	
	The EOS should therefore be used to determine the relation between $\mathcal{H}(T)$ and $\Theta(T)/T^4$, rather than assuming that they are identical.
	
	For numerical applications, $\mathcal{H}(T)$ can be evaluated directly from lattice-QCD parameterizations of $\epsilon(T)$ and $p(T)$. Moreover, since its definition involves first derivatives, its numerical determination should include uncertainty propagation from the EOS and a stability analysis of the interpolation procedure. This naturally motivates a comparison with related thermodynamic observables: in particular, both the location and height of any maximum in $\mathcal{H}(T)$ should be compared with those of established crossover indicators, including the interaction measure, the chiral susceptibility, and the minimum of the speed of sound, as mentioned earlier.
	
	\subsection{Relation to Other Thermodynamic Observables}
	
	The quantity \(\mathcal{H}(T)\) is evaluated consistently with the other thermodynamic observables obtained from the same EOS. For compactness, we resume these quantities as the vector
	\begin{equation}\label{eq:nadademais}
		\mathbf{X}(T) =
		\left[
		\widetilde{w}(T),
		\mathcal{H}(T),
		\frac{\Theta(T)}{T^4},
		c_s^2(T)
		\right],
	\end{equation}
	used solely as an organizational device for a set of complementary diagnostics of the EOS. It does not imply the introduction of a thermodynamic metric, a tensorial structure, or any additional geometric construction. Actually, each component of \(\mathbf{X}(T)\) emphasizes a different aspect of the system: \(\widetilde{w}(T)\) characterizes the enthalpy density, \(\mathcal{H}(T)\) measures its logarithmic thermal variation, \(\Theta(T)/T^4\) quantifies the breaking of conformal symmetry, and \(c_s^2(T)\) measures the stiffness of the medium.
	
	Since these quantities are derived from the same thermodynamically consistent EOS, they are linked by exact identities rather than independent inputs. This underlies the identity derived in the next subsection, connecting \(\mathcal{H}(T)\) and \(c_s^2(T)\) within a common thermodynamic framework. However, their distinct physical meanings mean their characteristic structures need not coincide, so their maxima or minima may occur at different temperatures. In particular, a maximum of \(\mathcal{H}(T)\) should not be taken as the maximum of the interaction measure, the minimum of \(c_s^2(T)\), or the chiral pseudocritical temperature.
	
	\subsection{Exact thermodynamic identity}
	\label{sec:thermodynamic_identity}
	
	Here, we stress that the quantities in equation \eqref{eq:nadademais} are not independent when they are calculated from the same EOS. To see this, one defines
	\begin{equation}
		I(T)\equiv\frac{\Theta(T)}{T^4},
	\end{equation}
	then, the normalized enthalpy given by equation \eqref{eq:dimensionless_enthalpy} can be written as
	\begin{equation}
		\widetilde{w}(T)  = 4\frac{p(T)}{T^4}+I(T),
	\end{equation}
	and since
	\begin{equation}
		\frac{d}{d\ln T}\left(\frac{p}{T^4}\right)=I(T),
	\end{equation}
	one obtains
	\begin{equation}
		\mathcal{H}(T)
		=
		\frac{4I(T)+dI(T)/d\ln T}{\widetilde{w}(T)}.
		\label{eq:H_trace_identity}
	\end{equation}
	
	Of course, the squared speed of sound satisfies the relation
	\begin{equation}\label{eq:cs_trace_identity}
		c_s^2(T)
		=
		\frac{\widetilde w(T)}
		{3\widetilde w(T)+4I(T)+dI(T)/d\ln T}
		=\frac{\widetilde w(T)}
		{d(\epsilon/T^4)/d\ln T+4\epsilon/T^4},
	\end{equation}
	and combining equations~\eqref{eq:H_trace_identity} and
	\eqref{eq:cs_trace_identity}, one writes
	\begin{equation}\label{eq:thermodynamic_identity}
		\mathcal H(T)=\frac{1}{c_s^2(T)}-3,
	\end{equation}
	providing an internal consistency relation among the quantities shown in equation \eqref{eq:nadademais}, clarifying that their temperature dependencies are related but {\it not identical}. 
	
	\section{Analytical Reference Equations of State}
	\label{sec:ref_eos}
	
	\subsection{MIT Bag Model}
	
	To examine the thermal response of a simple analytical EOS, we consider the simple MIT Bag Model. For a gas of massless, non-interacting quarks and gluons, the energy density and pressure are given by \cite{chodos1974,shuryak1980}
	\begin{equation}\label{eq:mit_eos}
		\epsilon(T)=\sigma T^4+\mathcal{B},
		\qquad
		p(T)=\frac{1}{3}\sigma T^4-\mathcal{B},
	\end{equation}
	where $\sigma=(g_{\mathrm{eff}}\pi^2)/30$ is the Stefan-Boltzmann coefficient and $\mathcal{B}$ is the bag constant, being a temperature-independent vacuum contribution to the energy density and pressure. The corresponding dimensionless thermodynamic quantities in this model are given by
	\begin{equation}\label{eq:dimensionless_mit}
		\widetilde{\epsilon}(T) \equiv
		\frac{\epsilon(T)}{T^4} = \sigma+\frac{\mathcal{B}}{T^4},
		\qquad
		\widetilde{p}(T) \equiv \frac{p(T)}{T^4}  =
		\frac{\sigma}{3}-\frac{\mathcal{B}}{T^4},
	\end{equation}
	and the resulting dimensionless enthalpy density is
	\begin{align}\label{eq:dimensionless_mit_enthalpy}
		\widetilde{w}(T)  &\equiv \frac{\epsilon(T)+p(T)}{T^4}=
		\widetilde{\epsilon}(T)+\widetilde{p}(T)
		=  \frac{4}{3}\sigma,
	\end{align}
	implying that the vacuum contributions cancel exactly in the enthalpy density. Then, the thermal response, defined in equation~\eqref{eq:enthalpy_response}, therefore becomes
	\begin{equation}
		\mathcal{H}_{\mathrm{MIT}}(T) \equiv \frac{d\ln\widetilde{w}(T)}{d\ln T}
		=
		\frac{d}{d\ln T} \ln\left(\frac{4}{3}\sigma\right)
		= 0.
		\label{eq:H_mit}
	\end{equation}
	
	Of course, the same result follows from the weighted representation, observing that in the physical domain where $\epsilon(T)+p(T)>0$, the thermodynamic weights are given by
	\begin{equation}\label{eq:mit_thermodynamic_weights}
		g(T) =  \frac{\epsilon(T)}{\epsilon(T)+p(T)}
		= \frac{3}{4}
		+ \frac{3\mathcal{B}}{4\sigma T^4},
		\qquad
		f(T) = \frac{p(T)}{\epsilon(T)+p(T)}
		= \frac{1}{4}  - \frac{3\mathcal{B}}{4\sigma T^4},
	\end{equation}
	and satisfy $f(T)+g(T)=1$. Since
	\begin{equation}
		\frac{d\ln\widetilde{\epsilon}}{d\ln T}
		= -\frac{4\mathcal{B}}  {\sigma T^4+\mathcal{B}},
		\qquad
		\frac{d\ln\widetilde{p}}{d\ln T} = \frac{4\mathcal{B}}
		{\sigma T^4/3-\mathcal{B}},
	\end{equation}
	their weighted sum vanishes identically, in agreement with
	equation~\eqref{eq:H_mit}. Actually, the vanishing of $\mathcal{H}_{\mathrm{MIT}}(T)$ does not imply that the MIT Bag Model is fully conformal. Indeed, its trace anomaly is given by
	\begin{equation}
		\Theta(T) \equiv \epsilon(T)-3p(T) = 4\mathcal{B},
	\end{equation}
	implying in
	\begin{equation}\label{eq:mit_trace_anomaly}
		\frac{\Theta(T)}{T^4} =
		\frac{4\mathcal{B}}{T^4},
	\end{equation}
	which is relevant since the trace anomaly quantifies the deviation from the conformal relation $\epsilon=3p$, whereas $\mathcal{H}(T)$ quantifies the temperature dependence of the dimensionless enthalpy density. We reinforce that these are distinct observables and need not have the same temperature dependence. The temperature at which the pressure vanishes is therefore \cite{chodos1974,shuryak1980}
	\begin{equation}\label{eq:mit_zero_pressure_temperature}
		T_0  = \left(\frac{3\mathcal{B}}{\sigma}\right)^{1/4}.
	\end{equation}
	
	For $T<T_0$, the simple deconfined-phase expression gives negative pressure and should not be treated as a physical description of the QCD medium. The apparent singularity of $d\ln \widetilde{p}/dT$ at $T=T_0$ is a mathematical artifact that arises from differentiating the logarithm at the point where the pressure vanishes. Since $p(T_0)=0$, the expression $d\ln p/dT = (1/p)(dp/dT)$ diverges; however, this behavior does not indicate a physical phase transition or an instability in the thermodynamic system, which remains well-defined near the bag-model vacuum temperature. As a result, the logarithmic derivative diverges even though the underlying thermal response does not.
	
	
	\subsection{Single-pion reference}
	
	For a classical relativistic gas of massive particles, the pressure and enthalpy are obtained from the standard thermodynamic relations $s=dp/dT$ and $w=\epsilon+p=Ts$. For a single pion species in the Boltzmann approximation, \begin{equation}
		p_\pi(T)=\frac{g_\pi m_\pi^2T^2}{2\pi^2} K_2\left(\frac{m_\pi}{T}\right),
	\end{equation}
	and the corresponding enthalpy is given by \cite{Cleymans:1999st}
	\begin{equation}
		w_\pi(T)= \frac{g_\pi m_\pi^2T^2}{2\pi^2} \left[4T K_2\left(\frac{m_\pi}{T}\right) +m_\pi K_1\left(\frac{m_\pi}{T}\right) \right].
	\end{equation}
	
	Consequently, the dimensionless ratio $(m_\pi/T)K_1(m_\pi/T)/K_2(m_\pi/T)$ measures the departure from the massless, conformal limit. We therefore use the single-pion reference as a qualitative benchmark for the thermal response,
	\begin{equation}\label{eq:spr}
		\mathcal{H}_{\mathrm{SPR}}(T) = \frac{m_\pi}{T} \frac{K_1(m_\pi/T)}{K_2(m_\pi/T)}-1.
	\end{equation}
	
	This reference is not meant to reproduce the full QCD EOS, but rather to isolate the characteristic temperature dependence produced by a massive relativistic excitation. In particular, \(\mathcal{H}_{\mathrm{SPR}}\) approaches its relativistic limiting behavior when \(T\gg m_\pi\) and grows in magnitude as the nonrelativistic regime \(T\ll m_\pi\) is approached, with its sign and size indicating the direction and degree of departure from conformal behavior.

	\section{Lattice Data and Numerical Method}
	\label{sec:lattice}
	
	\subsection{Wuppertal--Budapest data}
	
	We apply equation~\eqref{eq:enthalpy_response} to continuum-extrapolated EOS data from the Wuppertal--Budapest Collaboration \cite{borsanyi2010}. The input data provide the temperature dependence of the interaction measure, equation \eqref{eq:dimensionless_trace_anomaly}, from which the normalized pressure, energy density, and enthalpy are obtained using the thermodynamic relation between the interaction measure and pressure. The normalized enthalpy is then given by equation \eqref{eq:dimensionless_enthalpy}. 
	
	The shaded regions in the figures correspond to the central $68\%$ confidence intervals, defined by the 16th and 84th percentiles of the Monte Carlo (MC) distributions. These intervals include propagated uncertainties in the input data and, when available, the correlations among them. They also depend on the adopted smoothing prescription.
	
	\subsection{Uncertainty propagation}
	
	Input uncertainties are propagated via MC sampling, and a covariance matrix is available. Each realization is drawn from a multivariate normal distribution with the lattice data as the mean and the total covariance matrix as the covariance. For each realization, the pressure and normalized enthalpy are reconstructed, and a cubic smoothing spline is fitted to $\widetilde{w}(T)$. The numerical analysis uses $N_{\rm MC}=3000$ MC realizations, a cubic smoothing spline with $k=3$ and $s=0.5$, and a dense temperature grid to evaluate derivatives and extrema. The thermal response is obtained from the analytical derivative of the fitted spline, given by equation \eqref{eq:enthalpy_response}.
	
	To place $\mathcal{H}(T)$ in the EOS context, we compare \(\mathcal{H}(T)\) with the normalized interaction measure, \(\Theta(T)/T^4\), and the squared speed of sound, \(c_s^2(T)\). These observables arise from the same thermodynamic input but probe different properties of the medium.
	
	\section{Results}
	\label{sec:resultados}
	
	\subsection{Normalized enthalpy and $\mathcal{H}(T)$}

\begin{figure}[t]
		\centering
		\includegraphics[
		width=0.7\linewidth]{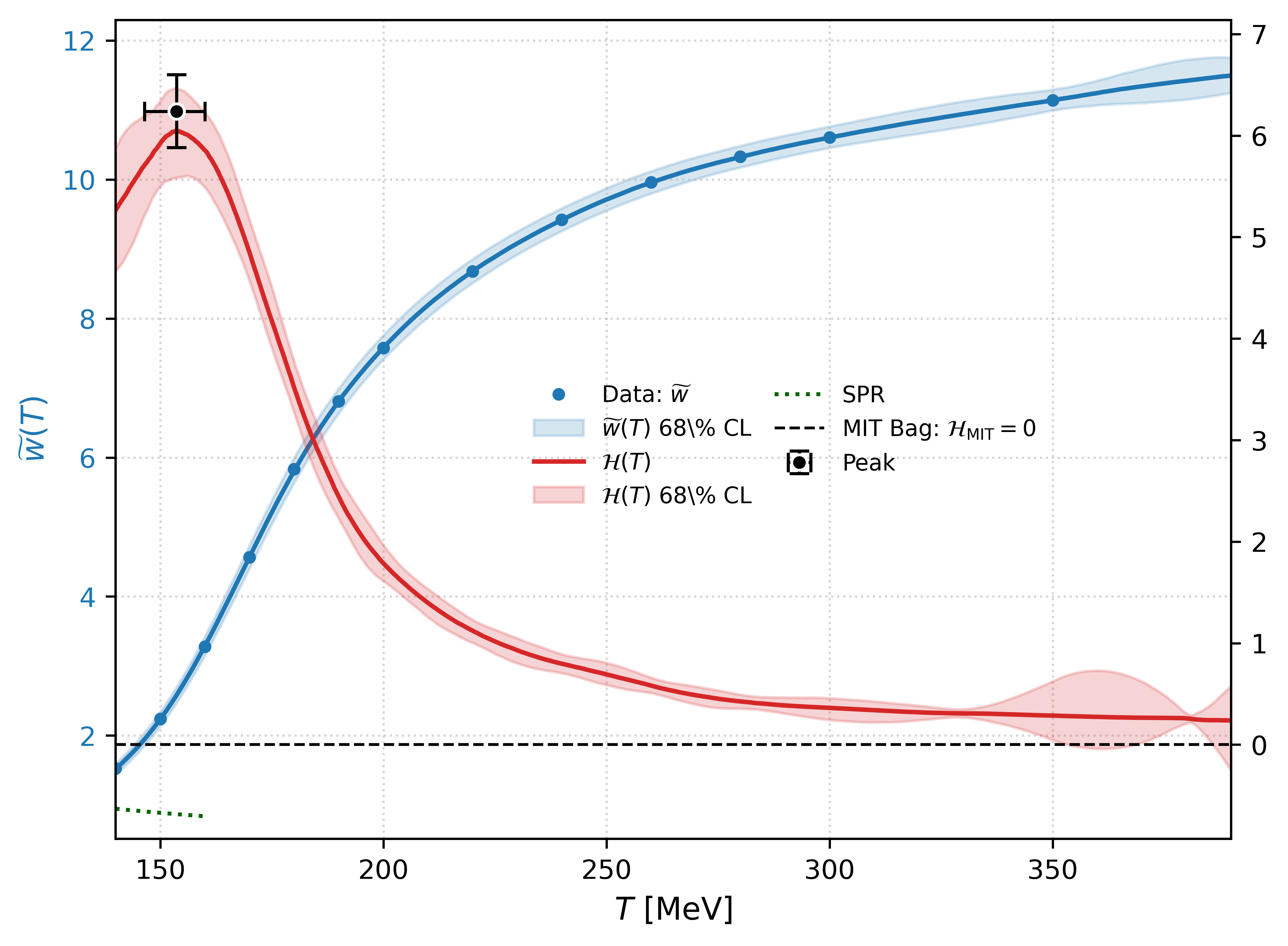}
		\caption{Normalized enthalpy density $\widetilde{w}(T)$ and $\mathcal{H}(T)$, obtained from the Wuppertal--Budapest lattice-QCD EOS. Blue points and the curve show $\widetilde{w}(T)$; the red curve shows the median response, and the shaded band gives the central $68\%$ confidence interval from MC propagation of input uncertainties. The dotted curve is the SPR of equation~\eqref{eq:spr}, included as a qualitative low-temperature comparison. The dashed line denotes the MIT Bag Model baseline, $\mathcal{H}_{\mathrm{MIT}}=0$. The maximum of the lattice thermal response is a finite indicator of rapid variation in normalized enthalpy, not evidence of a singularity or universal critical scaling. }
		\label{fig:h_comparison}
	\end{figure}

	Figure~\ref{fig:h_comparison} shows the normalized enthalpy and $\mathcal{H}(T)$. The normalized enthalpy increases with temperature and exhibits a crossover enhancement over an intermediate temperature range before becoming smoother at high temperatures. This trend reflects the gradual activation of thermodynamic degrees of freedom across the crossover region \cite{bazavov2014}.
	
	The thermal response shows a broad enhancement in the crossover region, which can be viewed as a signature of rapid changes in the effective thermodynamic degrees of freedom as the system evolves through the transition. This enhancement is followed by a decline toward smaller values at higher temperatures, consistent with the approach to approximate conformal scaling, where $\widetilde{w}(T)$ varies slowly. Within the uncertainty band, however, the data do not establish strict monotonicity in the high-temperature tail.
	
	At low temperature, the SPR, given by equation \eqref{eq:spr}, encodes the qualitative behavior expected from a massive hadronic excitation \cite{Cleymans:1999st,tawfik2005}. It is important to stress that its role here is limited to providing a reference scale for the low-temperature thermal response, providing a natural baseline for the MC peak estimate that follows. The lattice thermal response departs from this simple hadronic behavior as additional thermodynamic degrees of freedom become relevant. Therefore, the SPR curve does not represent a complete quantum-statistical Hadron Resonance Gas calculation. 
	
	The peak obtained from the MC realizations is
	\begin{equation}
		T_{\rm peak}
		=
		153.64^{+6.38}_{-7.13},
		\qquad
		\mathcal{H}_{\rm peak}
		= 6.24^{+0.36}_{-0.36}
	\end{equation}
	where the quoted interval corresponds to the central $68\%$ confidence range for the peak temperature. It is important to stress that this peak should be regarded as a property of the smoothed, uncertainty-propagated EOS representation, so its numerical value may vary with the temperature range, input covariance, and smoothing prescription. In the MC analysis, we varied the smoothing prescription and covariance choices across sampled realizations to assess their impact on the inferred peak position.
	
	However, Table~\ref{tab:table_1} shows that the peak sensitivity test in $\mathcal{H}(T)$ depends only weakly on the smoothing parameter $s$ used for interpolation by \texttt{UnivariateSpline}. For the three values analyzed, $s=0.1$, $0.5$, and $1.0$, the peak at the median temperature remains practically unchanged, varying from approximately $153.3$ to $153.8$ MeV, indicating that the location of $T_{\rm peak}$ is robust against reasonable changes in the smoothing choice and is therefore unlikely to be an artifact of the interpolation procedure. Moreover, the shift is much smaller than the width of the two MC confidence intervals, suggesting that the uncertainty due to the smoothing parameter is negligible compared with the statistical uncertainty.
	\begin{table}[ht]
		\centering
		\begin{tabular}{c|ccc|ccc}
			\hline
			$s$ &
			\multicolumn{3}{c|}{$T_{\rm peak}$ (MeV)} &
			\multicolumn{3}{c}{$\mathcal{H}_{\rm peak}$} \\
			\cline{2-7}
			&
			median & low & high &
			median & low & high \\
			\hline\hline
			0.1 & 153.76 & 145.26 & 161.02 & 6.26 & 5.88 & 6.66 \\
			\hline
			0.5 & 153.63 & 146.51 & 160.02 & 6.24 & 5.88 & 6.60 \\
			\hline
			1.0 & 153.26 & 146.72 & 157.02 & 6.25 & 5.89 & 6.60 \\
			\hline
		\end{tabular}
		\caption{Stability of the peak position and peak amplitude of $\mathcal{H}(T)$ under variations of the smoothing parameter $s$ used in cubic \texttt{UnivariateSpline} interpolation. The median, low, and high entries correspond to the 50th, 16th, and 84th percentiles, respectively, defining an approximately 68\% confidence interval. $T_{\rm peak}$ is expressed in MeV, while $\mathcal{H}_{\rm peak}$ is dimensionless. The weak dependence on $s$ shows the robustness of the extracted peak.}
		\label{tab:table_1}
	\end{table}
	
	The peak amplitude is similarly stable for $\mathcal{H}_{\rm peak}$ varying only between about $6.24$ and $6.26$, a relative change of less than $0.3\%$. This smoothing-induced spread is small compared with the widths of the approximately $68\%$ MC confidence intervals, which substantially overlap across different values of $s$. Thus, the peak-amplitude estimate is not significantly affected by the smoothing parameter $s$ within the considered interval.
	
	The small differences between the results can be attributed to smoothing in the numerical derivatives and the propagation of statistical fluctuations from the MC realizations. In particular, increasing $s$ from $0.1$ to $1.0$ slightly lowers the upper bound of the 68$\%$ confidence interval for $T_{\rm peak}$, but leaves its central value essentially unchanged.
	
	Therefore, the results show that identifying the peak of $\mathcal{H}(T)$ is numerically robust for $0.1\leq s\leq1.0$, supporting the interpretation of the peak as a feature of the analyzed EOS rather than an artifact of a particular interpolation procedure. However, this verification only tests numerical robustness; it does not by itself demonstrate universality or critical behavior in $\mathcal{H}(T)$. 
	\begin{figure}[h!]
		\centering
		\includegraphics[
		width=0.7\linewidth
		]{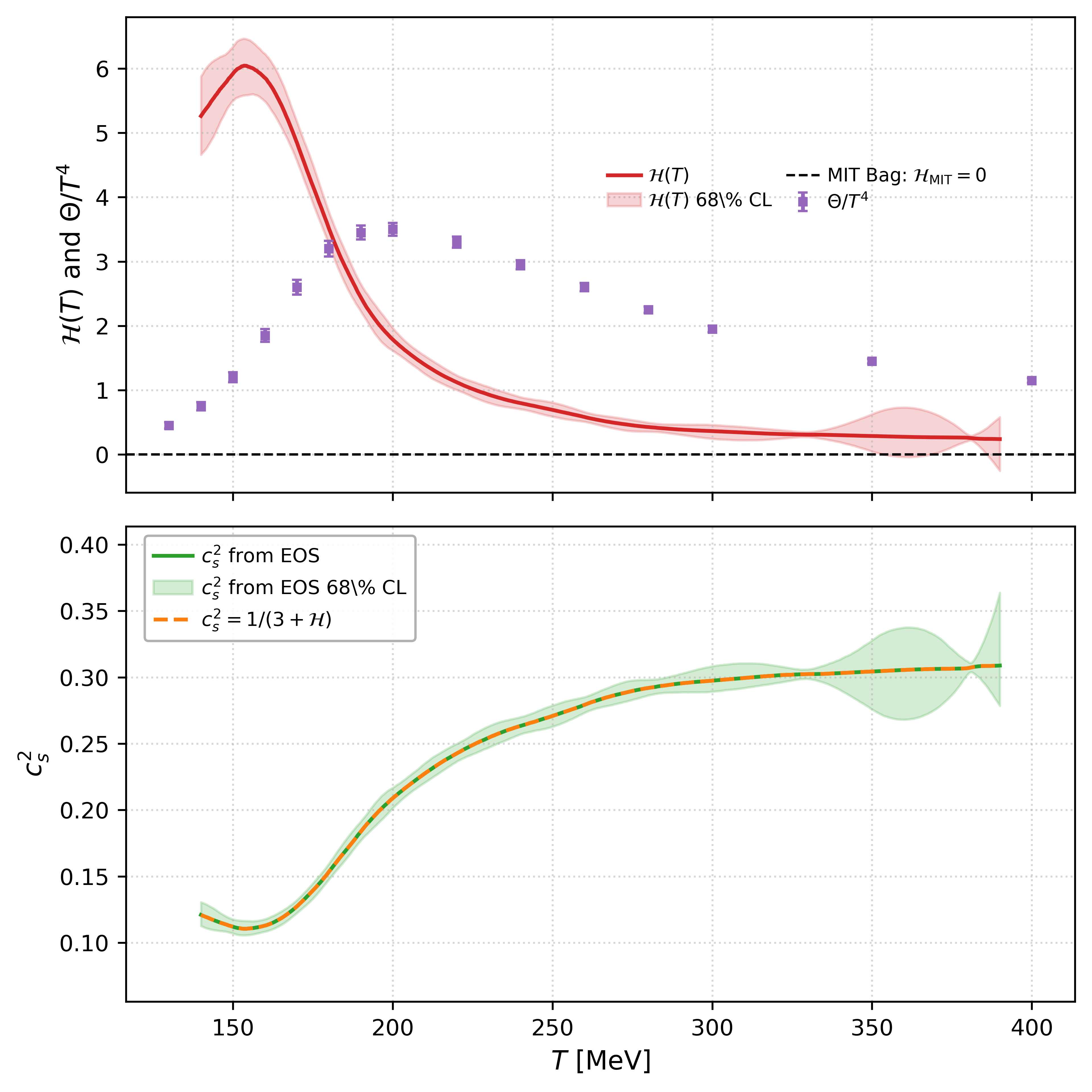}
		\caption{Upper panel: \(\mathcal{H}(T)\), together with the 68\% confidence interval, the MIT Bag Model reference \(\mathcal{H}{\mathrm{MIT}}=0\), and the trace anomaly \(\Theta/T^4\). Lower panel: squared speed of sound obtained directly from the EOS, \(c_{s,\mathrm{EOS}}^2(T)\), compared with the thermodynamic reconstruction \(c_{s,\mathrm{identity}}^2(T)\). The agreement between the two curves shows numerical thermodynamic consistency.}
		\label{fig:critical_scaling}
	\end{figure}
	
	\subsection{Comparison with $\Theta/T^4$ and $c_s^2$}
	
	It is important to stress that the comparison performed here provides a consistency check for the thermodynamic interpretation of $\mathcal{H}(T)$ and should not be interpreted as a universal scaling plot. In particular, the observed enhancement of $\mathcal{H}(T)$ is not sufficient to determine a critical exponent or to establish the existence of a critical endpoint. Since $\mathcal{H}(T)$, $\Theta/T^4$, and $c_s^2$ probe different thermodynamic derivatives and combinations thereof, {\it they need not peak or dip at exactly the same temperature}.
	
	In the upper panel of  Figure~\ref{fig:critical_scaling}, the thermal response \(\mathcal{H}(T)\) and the interaction measure \(\Theta(T)/T^4\) exhibit distinct temperature dependences, and their extrema need not coincide. The lower panel shows that \(c_s^2(T)\) obtained directly from the EOS agrees with the reconstruction \(1/[3+\mathcal{H}(T)]\) throughout the temperature interval, providing a numerical consistency check of the thermodynamic reconstruction rather than evidence for an independent physical comparison. Accordingly, the comparison between $\mathcal{H}(T)$ and $c_s^2(T)$ should be interpreted as a consistency test of the numerical reconstruction rather than as a comparison between independent thermodynamic observables. 
	
\begin{figure}[h!]
		\centering
		\includegraphics[width=0.7\linewidth]{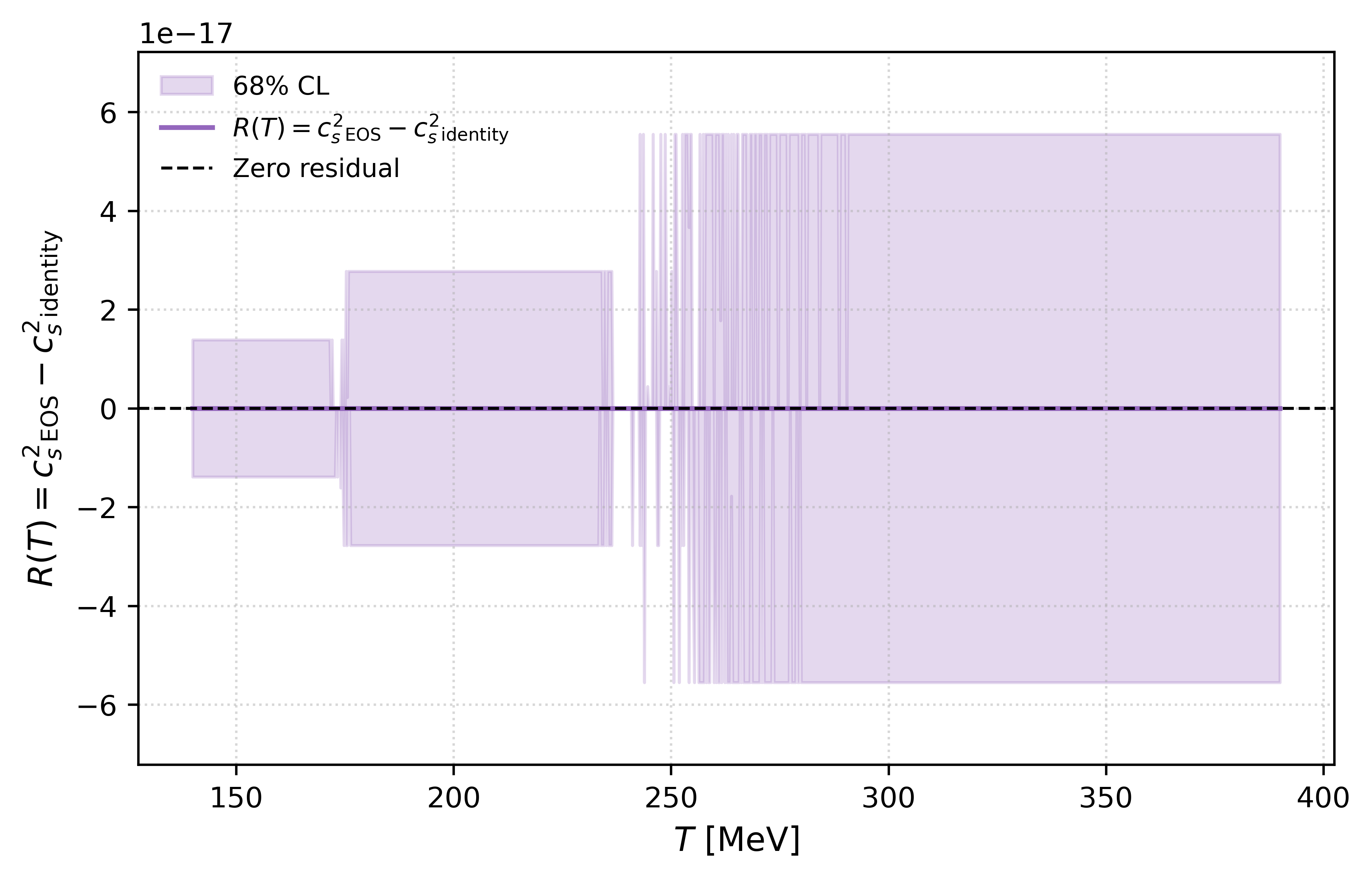}
		\caption{Numerical consistency residual of the thermodynamic identity as a function of temperature, \(R(T)\). Here, \(c_{s,\mathrm{EOS}}^2\) is obtained from the reconstructed EOS, whereas \(c_{s,\mathrm{identity}}^2\), equation~\eqref{eq:csidentity}, follows from the exact relation \(\mathcal{H}(T)\), equation~\eqref{eq:thermodynamic_identity}. The solid curve shows the central value, the shaded region represents the 68\% confidence interval, and the dashed line shows the zero residual. The residual is consistent with zero across the full temperature range, confirming the numerical thermodynamic consistency of the reconstruction.}
		\label{fig:soundspeed_residual}
	\end{figure}

	Actually, in the present implementation, the residual can be defined as
	\begin{eqnarray}
		R(T)=c_s^2(T)-c_{s,\mathrm{identity}}^2(T),
	\end{eqnarray}
	and since the underlying thermodynamic identity is exact, this agreement should not be interpreted as a comparison between independent physical observables. Rather, it provides a numerical consistency check of the EOS reconstruction when \(\mathcal{H}(T)\) and \(c_s^2(T)\) are calculated from the same MC realization and the same interpolated EOS. In this case, one writes
	\begin{equation}
		R(T)=c_{s,\mathrm{EOS}}^2(T)-c_{s,\mathrm{identity}}^2(T),
	\end{equation}
	where \(c_{s,\mathrm{EOS}}^2\) is obtained from the derivative of the reconstructed enthalpy density and
	\begin{equation}\label{eq:csidentity}
		c_{s,\mathrm{identity}}^2(T)
		= \frac{1}{3+\mathcal{H}(T)}
	\end{equation}
	is calculated from the exact thermodynamic identity given by equation~\eqref{eq:thermodynamic_identity}. Figure~\ref{fig:soundspeed_residual} shows that the residual vanishes within floating-point precision, remaining indistinguishable from zero over the entire temperature range, with a magnitude of approximately \(10^{-17}\), i.e., at the level of floating-point and interpolation precision. The central curve is therefore fully consistent with the zero-residual line, while the narrow uncertainty band remains centered around zero.
	
	From Figure \ref{fig:h_comparison}, one can see this result shows the internal thermodynamic consistency of the reconstruction procedure and, in particular, it shows that \(\mathcal{H}(T)\) and \(c_s^2(T)\), evaluated from the same MC realization and the same interpolant of \(\widetilde{w}(T)\), satisfy the exact thermodynamic identity within numerical precision. However, we stress that the residual should therefore be interpreted as a numerical consistency diagnostic, not as an independent physical observable or evidence of a thermodynamic violation.

	\section{Interpretation and Limitations}
	\label{sec:limitations}
	
	\subsection{High-temperature behavior}
	
	As aforementioned, at high temperature, QCD approaches an approximately conformal regime. In this limit, the EOS approaches the Stefan--Boltzmann behavior,
	\begin{equation}
		\epsilon(T) \simeq 3p(T) \propto T^4,
	\end{equation}
	and consequently, the normalized enthalpy density given by equation \eqref{eq:dimensionless_enthalpy} approaches a slowly varying function and, asymptotically, a constant ideal-gas value. From the definition of the thermal response given by equation~\eqref{eq:enthalpy_response}, it follows that a temperature-independent normalized enthalpy satisfies
	\begin{equation}
		\lim_{T\rightarrow\infty}\mathcal{H}(T)=0.
	\end{equation}
	
	The central lattice estimate shows a decrease of $\mathcal{H}(T)$ above the
	crossover region and approaches relatively small values at high temperature.
	Despite this result, the propagated uncertainty band does not support a strong claim of strict monotonicity in the high-temperature tail. However, the observed trend is nevertheless compatible with the approach to approximate conformal behavior.
	
	This result differs from the MIT Bag Model, for which
	$\mathcal{H}_{\mathrm{MIT}}(T)=0$ at every temperature by construction. In
	that model, the temperature-independent vacuum contribution cancels from the
	enthalpy density given by equation \eqref{eq:enthalpy_density}. The lattice-QCD result is instead sensitive to residual non-conformal effects, including the running coupling, finite quark masses, and nonperturbative dynamics.
	
	\subsection{Crossover interpretation}
	
	The thermal response is evaluated using continuum-extrapolated
	lattice-QCD equations of state, which provide non-perturbative
	determinations of the pressure $p(T)$ and energy density $\epsilon(T)$
	across the QCD crossover region
	\cite{bazavov2014,borsanyi2010,borsanyi2020}. Defining the normalized through equation \eqref{eq:dimensionless_enthalpy}, then the thermal response is given by 
	\begin{equation}
		\mathcal{H}(T)
		\equiv
		\frac{d\ln\widetilde{w}(T)}{d\ln T}
		=
		T\frac{d}{dT}
		\ln\left[
		\frac{\epsilon(T)+p(T)}{T^4}
		\right].
		\label{eq:H_lattice_definition}
	\end{equation}
	
	In practice, lattice-QCD data are represented by a smooth interpolation of the EOS over the temperature interval of interest. As aforementioned, $\mathcal{H}(T)$ involves a derivative, implying its numerical determination is sensitive to the interpolation procedure and the uncertainties in the underlying EOS. The robustness of the result should therefore be assessed with alternative interpolation or smoothing prescriptions, together with appropriate propagation of lattice uncertainties. For a continuum-extrapolated EOS, the dataset, parametrization, temperature range, and uncertainty treatment must be specified explicitly \cite{borsanyi2010,borsanyi2020}.
	
	The resulting thermal response can be compared with other thermodynamic characterizations of the crossover, including the interaction measure (defined in equation~\eqref{eq:dimensionless_trace_anomaly}), the chiral susceptibility, and the squared speed of sound. A localized maximum or rapid variation of $\mathcal{H}(T)$ would indicate a change in the scaling behavior of the normalized enthalpy. However, its location need not coincide with the maximum of any other crossover observable. Such comparisons should therefore be performed numerically, accounting for uncertainties and the different definitions of the pseudocritical temperature.
	
	At high temperatures, approximate conformal behavior implies that dimensionless thermodynamic quantities vary slowly. In the ideal conformal limit, $\widetilde{w}(T)$ is constant and $\mathcal{H}(T)\longrightarrow 0$. 
	For QCD at physically relevant temperatures, this limit is reached only gradually because scale invariance is broken by the running coupling and finite quark masses. Thus, the asymptotic behavior of $\mathcal{H}(T)$ should be viewed as an approach to conformal scaling rather than its exact realization.
	
	\subsection{Critical-scaling limitations}
	
	
	The thermal response defined in equation~\eqref{eq:enthalpy_response} is constructed from the full EOS, and it is not itself a critical exponent, an order-parameter susceptibility, or a universal scaling function. Near a genuine critical point, the normalized enthalpy may contain regular and
	singular contributions \cite{stephanov2004,parotto2020}
	\begin{equation}
		\widetilde{w}(T,\mu_B) =
		\widetilde{w}_{\mathrm{reg}}(T,\mu_B) +
		\widetilde{w}_{\mathrm{sing}}(T,\mu_B).
	\end{equation}
	where these contributions to $\mathcal{H}(T)$ depend on the relevant
	scaling fields, the regular background, and the trajectory through the
	$(T,\mu_B)$ phase diagram. Consequently, the critical exponent $\alpha$ alone
	does not determine the height, width, or position of a maximum in
	$\mathcal{H}(T)$.
	
	In the chiral limit, two-flavor QCD is expected to belong to the three-dimensional $O(4)$ universality class \cite{pisarski1984}, where the critical exponent $\alpha$ is negative ($\alpha \approx -0.21$ in 3D), implying that the heat capacity does not diverge at the transition point, but exhibits a cusp-like singularity instead \cite{Pelissetto2002}. At finite baryon density, a possible critical endpoint is commonly associated with 3D Ising-like behavior \cite{stephanov2004}. Applying these scaling arguments to QCD requires, however, a specified mapping between QCD variables and scaling fields, a treatment of the regular background, and information along trajectories at nonzero baryon chemical potential \cite{parotto2020}.
	
	In QCD at physical quark masses and $\mu_B=0$, the transition from a hadron-dominated medium to a QGP medium is a smooth crossover rather than a genuine thermodynamic phase transition \cite{aoki.2006,borsanyi2010}. Consequently, thermodynamic observables evolve continuously over a finite temperature interval, although their susceptibilities and response functions may exhibit pronounced but finite maxima. Accordingly, the finite enhancement in $\mathcal{H}(T)$ should be interpreted as a thermodynamic indicator of a rapid change in the normalized enthalpy. 
	
	More generally, the temperature dependence of $\mathcal{H}(T,\mu_B)$ may be a useful diagnostic of EOS changes: A sharper $\mathcal{H}(T)$ with increasing $\mu_B$ could be consistent with approaching a critical region, but it would not, by itself, determine an Ising critical exponent or establish a critical endpoint. 
	
	\section{Concluding Remarks and Prospects}
	\label{sec:final_remarks}
	
	We have introduced the thermal response $\mathcal{H}(T)$, defined as the logarithmic temperature variation of the normalized enthalpy density. At physical quark masses and zero baryon chemical potential, the maximum of $\mathcal{H}(T)$ serves as a diagnostic of rapid changes in the active thermodynamic degrees of freedom across the QCD crossover, rather than as a signal of a true critical singularity.
	
	Through the exact thermodynamic identity \(\mathcal{H}(T)=1/c_s^2(T)-3\), \(\mathcal{H}(T)\) is directly related to the squared speed of sound. It adds no independent thermodynamic information beyond \(c_s^2(T)\), but serves as a sensitive visual and analytical diagnostic: large \(\mathcal{H}(T)\) values can distinguish the EOS softest point, where \(c_s^2\) is suppressed. Moreover, in the high-temperature limit of an ideal conformal gas, \(\mathcal{H}(T)\to 0\) as \(c_s^2\to 1/3\). The algebraic equivalence of \(c_s^2(T)\) and \(\mathcal{H}(T)\) also provides a rigorous consistency check for numerical EOS reconstructions, validated to floating-point precision (\(\sim 10^{-17}\)). Differences between \(\mathcal{H}(T)\) and the trace anomaly \(\Theta(T)/T^4\) reflect the distinct thermodynamic combinations each quantity samples.
	
	Of course, a full determination of universal critical scaling would require additional input, including light-quark mass dependencies, explicit scaling-field mappings, and a controlled subtraction of the regular thermodynamic background. Likewise, a Ruppeiner-geometric analysis \cite{rup1995} requires an EOS dependent on at least two independent thermodynamic variables and is beyond the scope of this one-dimensional study.
	
	From a phenomenological perspective, evaluating $\mathcal{H}(T)$ with continuum-extrapolated lattice QCD data produces a clear peak at $T_{\text{peak}} = 153.64_{-7.13}^{+6.38}\text{ MeV}$, with amplitude $\mathcal{H}_{\text{peak}} = 6.24 \pm 0.36$. Future work will extend this framework to alternative continuum-extrapolated lattice parameterizations and smoothing methods. Crucially, extending this diagnostic to finite baryon chemical potential ($\mu_B > 0$) offers a promising path: if $\mathcal{H}(T, \mu_B)$ sharpens along the chemical freeze-out line, it could serve as a highly sensitive marker for locating the hypothetical QCD critical point.

	\section*{Acknowledgments}
	
	SDC thanks UFSCar for its financial support.

	\bibliographystyle{unsrt}
	\bibliography{biblio}

@article{S.Weinberg.Phys.Rev.D8.3497.1973,
	author = {S. Weinberg},
	title = {New {A}pproach to the {R}enormalization {G}roup},
	journal = {Phys. Rev. D},
	volume = {8},
	pages = {3497},
	year = {(1973)}
}

@article{FuPawlowski2020,
	title = {{QCD} {P}hase {S}tructure at {F}inite {T}emperature and {D}ensity},
	author = {W.-J. Fu and J. M. Pawlowski and F. Rennecke},
	journal = {Phys. Rev. D},
	volume = {101},
	issue = {5},
	pages = {054032},
	numpages = {58},
	year = {(2020)},
	publisher = {American Physical Society},
	doi = {10.1103/PhysRevD.101.054032},
	url = {https://link.aps.org/doi/10.1103/PhysRevD.101.054032}
}

@article{borsanyi2010,
title={The {QCD} {E}quation of {S}tate with {D}ynamical {Q}uarks},
author={S. Borsanyi {\it et al.}},
journal={J. High Energ. Phys.},
volume={2010},
number={11},
pages={1--31},
year={(2010)},
publisher={Springer},
doi={10.1007/JHEP11(2010)077}
}

@article{shuryak1980,
title={The {P}hase {T}ransition in {Q}uantum {C}hromodynamics},
author={E. V. Shuryak},
journal={Phys. Rept.},
volume={61},
number={2},
pages={71--158},
year={(1980)},
publisher={Elsevier},
doi={10.1016/0370-1573(80)90105-2}
}

@article{bazavov2014,
title = {Equation of {S}tate in ($2+1$)-{F}lavor {QCD}},
author = {A. Bazavov {\it et al.} (HotQCD Collaboration)},
collaboration = {HotQCD Collaboration},
journal = {Phys. Rev. D},
volume = {90},
issue = {9},
pages = {094503},
numpages = {25},
year = {(2014)},
publisher = {American Physical Society},
doi = {10.1103/PhysRevD.90.094503},
url = {https://link.aps.org/doi/10.1103/PhysRevD.90.094503}
}

@book{callen1985,
title={Thermodynamics and an Introduction to Thermostatistics},
author={H. B. Callen},
edition={2nd},
year={(1985)},
publisher={John Wiley \& Sons},
address={New York}
}

@article{Pelissetto2002,
author = {A. Pelissetto and E. Vicari},
title = {Critical {P}henomena and {R}enormalization-{G}roup {T}heory},
journal = {Phys. Rept.},
volume = {368},
pages = {549-727},
year = {(2002)}
}

@article{Dupuis2021,
title = {The {N}onperturbative {F}unctional {R}enormalization {G}roup and its {A}pplications},
journal = {Phys. Rept.},
volume = {910},
pages = {1-114},
year = {(2021)},
issn = {0370-1573},
doi = {https://doi.org/10.1016/j.physrep.2021.01.001},
url = {https://www.sciencedirect.com/science/article/pii/S0370157321000156},
author = {N. Dupuis {\it et al.}}
}

@article{rup1995,
title={Riemannian {G}eometry in {T}hermodynamic {F}luctuation {T}heory},
author={G. Ruppeiner},
journal={Rev. Mod. Phys.},
volume={67},
number={3},
pages={605},
year={(1995)},
publisher={APS}
}

@article{karsch2002,
author = {F. Karsch},
title = {Lattice {R}esults on {QCD} {T}hermodynamics},
journal = {Nucl. Phys. A},
volume = {698},
pages = {199c-208c},
year = {(2002)}
}

@article{cheng2008,
author = {M. Cheng {\it et al.}},
title = {The {QCD} {E}quation of {S}tate with {A}lmost {P}hysical {Q}uark {M}asses},
journal = {Phys. Rev. D},
volume = {77},
pages = {014511},
year = {(2008)}
}

@article{ratti2018,
author = {C. Ratti},
title = {Lattice {QCD} and {H}eavy {I}on {C}ollisions: {A} {R}eview of {R}ecent {P}rogress},
journal = {Rept. on Prog. in Phys.},
volume = {81},
issue = {8},
pages = {084301},
year = {(2018)}
}

@article{chodos1974,
title = {New {E}xtended {M}odel of {H}adrons},
author = {A. Chodos {\it et al.}},
journal = {Phys. Rev. D},
volume = {9},
issue = {12},
pages = {3471--3495},
numpages = {0},
year = {(1974)},
month = {Jun},
publisher = {American Physical Society},
doi = {10.1103/PhysRevD.9.3471},
url = {https://link.aps.org/doi/10.1103/PhysRevD.9.3471}
}

@article{borsanyi2020,
author = {S. Borsanyi {\it et al.}},
title = {{QCD} {C}rossover at {F}inite {C}hemical {P}otential from {L}attice {S}imulations},
journal = {Phys. Rev. Lett.},
volume = {125},
pages = {052001},
year = {(2020)}
}

@article{wilson1974,
title = {The {R}enormalization {G}roup and the $\epsilon$ {E}xpansion},
journal = {Phys. Rept.},
volume = {12},
number = {2},
pages = {75-199},
year = {(1974)},
issn = {0370-1573},
doi = {https://doi.org/10.1016/0370-1573(74)90023-4},
url = {https://www.sciencedirect.com/science/article/pii/0370157374900234},
author = {K. G. Wilson and J. Kogut}
}

@article{borsanyi2025phase,
title={The {QCD} {P}hase {D}iagram},
author={S. Borsanyi and P. Parotto},
journal={arXiv:2512.08843},
year={(2025)}
}

@article{koch2025exploring,
title={Exploring the {QCD} {P}hase {D}iagram {T}hrough {C}orrelations and {F}luctuations},
author={V. Koch and V. Vovchenko},
journal={Eur. Phys. J. Spec. Top.},
year={(2025)}
}

@inbook{braun-munzinger2004,
author = {P. Braun-Munzinger and K. Redlich and J. Stachel},
title = {Particle {P}roduction in {H}eavy {I}on {C}ollisions},
booktitle = {Quark–Gluon Plasma 3},
chapter = {},
pages = {491-599},
year = {(2004)},
publisher = {World Scientific},
doi = {10.1142/9789812795533_0008},
URL = {https://www.worldscientific.com/doi/abs/10.1142/9789812795533_0008},
eprint = {https://www.worldscientific.com/doi/pdf/10.1142/9789812795533_0008}
}

@article{pisarski1984,
author    = {R. D. Pisarski and F. Wilczek},
title     = {Remarks on the {C}hiral {P}hase {T}ransition in {C}hromodynamics},
journal   = {Phys. Rev. D},
volume    = {29},
number    = {2},
pages     = {338-341},
year      = {(1984)},
doi       = {10.1103/PhysRevD.29.338}
}

@article{bazavov2019,
author = {A. Bazavov {\it et al.} (HotQCD Collaboration)},
title = {Chiral {C}rossover in {QCD} at {Z}ero and {N}on-{Z}ero {C}hemical {P}otentials},
journal = {Phys. Lett. B},
volume = {795},
pages = {15},
year = {(2019)}
}

@article{aoki.2006,
author = {Y. Aoki {\it et al.}},
title = {The {O}rder of the {Q}uantum {C}hromodynamics {T}ransition {P}redicted by the {S}tandard {m}odel of {P}article {P}hysics},
journal = {Nature},
volume = {443},
pages = {675-678},
year = {(2006)}
}

@article{aarts2016,
author = {G. Aarts},
title = {Introductory {L}ectures on {L}attice {QCD} at {N}onzero {B}aryon {N}umber},
journal = {J. Phys.: Conf. Ser.},
volume = {706},
pages = {022004},
year = {(2016)}
}

@article{petreczky2012,
author = {P. Petreczky},
title = {Lattice {QCD} at {N}on-{Z}ero {T}emperature},
journal = {J. of Phys. G: Nucl. and Part. Phys.},
volume = {39},
pages = { 093002},
year = {(2012)}
}

@article{Cleymans:1999st,
author  = {J. Cleymans and K. Redlich},
title   = {Unified {D}escription of {F}reeze-out {P}arameters in {R}elativistic {H}eavy-{I}on {C}ollisions},
journal = {Phys. Rev. C},
volume  = {60},
pages   = {054908},
year    = {(1999)},
doi     = {10.1103/PhysRevC.60.054908}
}

@article{tawfik2005,
author = {A. Tawfik},
title = {The {QCD} {P}hase {D}iagram: {A} {C}omparison of {L}attice and {H}adron {R}esonance {G}as {M}odel {C}alculations},
journal = {Phys.Rev. D},
volume = {71},
pages = {054502},
year = {(2005)}
}

@article{bazavov2009,
author = {A. Bazavov {\it et al.}},
title = {{E}quation of {S}tate and {QCD} {T}ransition at {F}inite {T}emperature},
journal = {Phys. Rev. D},
volume = {80},
pages = {014504},
year = {(2009)}
}

@article{bazavov2012,
author = {A. Bazavov {\it et al.} (HotQCD Collaboration)},
title = {{C}hiral and {D}econfinement {A}spects of the {QCD} {T}ransition},
journal = {Phys. Rev. D},
volume = {85},
pages = {054503},
year = {(2012)}
}

@article{parotto2020,
author    = {P. Parotto {\it et al.}},
title     = {{QCD} {E}quation of {S}tate {M}atched to the 3{D} {I}sing {U}niversality {C}lass},
journal   = {Phys. Rev. C},
volume    = {101},
number    = {3},
pages     = {034901},
year      = {(2020)},
doi       = {10.1103/PhysRevC.101.034901}
}

@article{stephanov2004,
author    = {M. A. Stephanov},
title     = {{QCD} {P}hase {D}iagram and the {C}ritical {P}oint},
journal   = {Prog. of Theor. Phys. Suppl.},
volume    = {153},
pages     = {139-156},
year      = {(2004)},
doi       = {10.1143/PTPS.153.139}
}
	
\end{document}